# Giant Spin Seebeck Effect through an Interface Organic Semiconductor

V. Kalappattil[1], R. Geng[2], R. Das[1], H. Luong[2], M. Pham[2], T. Nguyen[2], A. Popescu[1], L.M. Woods[1], M. Kläui[3], H. Srikanth[1], and M.H. Phan[1]

[1] Department of Physics, University of South Florida, Tampa, Florida 33620, USA

[2] Department of Physics and Astronomy, University of Georgia, Athens, GA 30602, USA

[3] Institute of Physics, Johannes Gutenberg University Mainz, 55128 Mainz, Germany

**Interfacing an organic semiconductor $C_{60}$ with a non-magnetic metallic thin film (Cu or Pt) has created a novel heterostructure that is ferromagnetic at ambient temperature, while its interface with a magnetic metal (Fe or Co) can tune the anisotropic magnetic surface property of the material. Here, we demonstrate that sandwiching $C_{60}$ in between a magnetic insulator ($Y_3Fe_5O_{12}$: YIG) and a non-magnetic, strong spin-orbit metal (Pt) promotes highly efficient spin current transport via the thermally driven spin Seebeck effect (SSE). Experiments and first principles calculations consistently show that the presence of $C_{60}$ reduces significantly the conductivity mismatch between YIG and Pt and the surface perpendicular magnetic anisotropy of YIG, giving rise to enhanced spin mixing conductance across YIG/$C_{60}$/Pt interfaces. As a result, a 600% increase in the SSE voltage ($V_{LSSE}$) has been realized in YIG/$C_{60}$/Pt relative to YIG/Pt. Temperature-dependent SSE voltage measurements on YIG/$C_{60}$/Pt with varying $C_{60}$ layer thicknesses also show an exponential increase in $V_{LSSE}$ at low temperatures below 200 K, resembling the temperature evolution of spin diffusion length of $C_{60}$. Our study emphasizes the important roles of the magnetic anisotropy and the spin diffusion length of the intermediate layer in** the SSE in YIG/C$_{60}$/Pt structures, providing a new pathway for developing novel spin-caloric materials.

Generating pure spin currents has been the main challenge for realizing highly efficient spintronics devices.[1] Notable effects under study for attaining pure spin currents are the spin Hall effect (SHE)[2], spin pumping mechanism[3,4], and the spin Seebeck effect (SSE).[5,6] In the last two approaches, a pure spin current is generated in a ferromagnetic (FM) material (which could be a metal[7], insulator[8], or semiconductor[9]) and converted into a voltage drop via the inverse spin Hall effect (ISHE) in a nonmagnetic metal (NM) possessing strong spin-orbit coupling. In case of the SSE, the spin current $J_s$ can be expressed as[10]

$$J_s = \frac{G}{2\pi} \frac{\gamma \hbar}{M_s V_a} K_b \Delta T, \qquad (1)$$

where $G$ is the spin mixing conductance, $\gamma$ is the gyromagnetic ratio, $V_a$ is the magnetic coherence volume, $M_s$ is the saturation magnetization, and $\Delta T$ is the temperature difference between magnons in FM and electrons in NM. One of the key parameters essential for an efficient spin transport across a FM/NM interface is having a large spin mixing conductance, which directly correlates with a large spin current, as evident from Eq. (1).[10] According to Eq. (1), increasing $G$ is essential to transport large spin currents from FM to NM.

Since the first observation of SSE in the ferromagnetic insulator Y$_3$Fe$_5$O$_2$ (YIG) in 2010,[8] this material has become one of the most intensively studied systems for fundamental understanding of the underlying spin transport and for prospective spin caloritronic applications, among others.[5,6] Significant efforts have been devoted towards improving the value of $G$ and hence the SSE voltage by utilizing YIG/Pt interfaces and adding various thin intermediate layers, such as Cu, NiO, CoO, Fe$_{70}$Cu$_{30}$, NiFe, or Py.[11-15] Inserting a non-magnetic layer of Cu, for example, between YIG and Pt has been reported to decrease the SSE signal[11], while the addition of an ultrathin magnetic layer of $Fe_{70}Cu_{30}$[14] or NiFe[15] between YIG and Pt has increased the magnetic moment density or $G$ at their interface, leading to an overall improvement in the SSE voltage. Recently, Lin *et al.* reported a significant enhancement of the SSE in YIG/*M*/Pt (*M* = NiO or CoO) heterostructures,[12] demonstrating the important role played by an additional antiferromagnetic layer on the spin current transport. Despite the enhancement of SSE reported in these heterostructures, the responsible underlying mechanisms associated with the role of the intermediate layer on the SSE have remained an open question.[10-15]

It is generally accepted that spin transport through a material is limited by its spin diffusion length $\lambda$. While YIG has a large $\lambda$ (~10 μm), a small $\lambda$ value of Pt (~2 nm) has been reported.[16] Consequently, undesired effects, such as large surface roughness of YIG[17] and/or large perpendicular surface magnetic anisotropy of YIG[18] causing strong spin scattering, could suppress considerably the spin current injection into the Pt layer. The large conductivity mismatch between YIG and Pt could also decrease considerably the efficient spin transport in YIG/Pt.[19] It is therefore desirable to seek an intermediate material that can reduce both the perpendicular surface magnetic anisotropy of YIG and the conductivity mismatch between the YIG and Pt layers.

In this regard, an organic semiconductor (OSC) such as $C_{60}$ buckyballs can be considered as a promising candidate material due to its low spin-orbit coupling that essentially results in weak spin scattering and consequently a large spin diffusion length ($\lambda$ can be of the order of several hundred nanometers).[20,21] Additionally, the $C_{60}$ buckyballs are semiconducting in nature [22], thus when sandwiched between YIG and Pt layers, the conductivity mismatch between them is likely to be reduced. Recent studies have also revealed that $C_{60}$ hybridizes strongly with metallic substrates, which leads to inducing ferromagnetic order at the surface layer of a non-magnetic metal, such as Cu or Pt[23], or tuning the surface magnetism (e.g. surface magnetic anisotropy) of a magnetic metal, such as Fe or Co.[24] These findings lead us to propose a new approach for improving the SSE in ferromagnetic insulator/metal systems such as YIG/Pt by adding a thin intermediate layer of an organic semiconductor such as $C_{60}$ and forming a novel heterostructure of YIG/$C_{60}$/Pt, as shown in Fig. 1. In this Letter, we synthesize YIG/$C_{60}$/Pt systems containing buckyball layers of various thicknesses and report the first comprehensive experimental demonstration of thermally generated pure spin currents in the systems through the longitudinal SSE (LSSE). Our experiments show that the spin current in the YIG/$C_{60}$/Pt heterostructures is significantly enhanced as compared to YIG/Pt structures. This enhancement can be as large as 600% and it is dependent on temperature and the thickness of the $C_{60}$ layer. We demonstrate that the long spin diffusion length of $C_{60}$ has a much pronounced positive impact on the SSE signal. The exponential temperature dependence of both the SSE voltage and $\lambda$ suggests that $\lambda$ scales with the SSE voltage in magnitude. Our first principles calculations based on density functional theory (DFT) corroborate the experimental measurements and provide further insight into the role of $C_{60}$ on the transport and magnetic properties of the YIG/$C_{60}$/Pt heterostructure.

**Effect of $C_{60}$ on the bulk and surface magnetic properties of YIG**: First, we show how the coating of $C_{60}$ modifies the bulk and surface magnetic properties of YIG by means of magnetometry and radio-frequency (RF) transverse susceptibility. Figure 2a shows the magnetic hysteresis (*M-H*) loops taken at 300 K for YIG, YIG/Pt and YIG/$C_{60}$(5 nm)/Pt structures. All three plots superimposed on each other, indicate that it is not possible to use magnetometry to identify the magnetic difference among these samples. In a recent study, we have demonstrated the excellent capacity of using our RF transverse susceptibility (TS) technique to probe the surface perpendicular magnetic anisotropy (PMA) field and its temperature evolution in YIG, providing the first experimental evidence for a strong effect of the surface PMA on $V_{\text{LSSE}}$.[25] Note that TS uses a self-resonant tunnel diode oscillator with a resonant frequency of ~12 MHz. The sensitivity of 10 Hz has been reached and validated by us over the years as a highly efficient tool for precisely measuring magnetic anisotropy in a wide range of magnetic materials.[25-29] In the present study, we have performed detailed TS experiments on YIG, YIG/Pt, and YIG/$C_{60}$/Pt structures. Figure 2b shows typical TS curves at 200 K for these samples as dc magnetic field was swept from positive saturation to negative saturation. As the field was swept from positive to negative saturation, the first peak corresponds to the bulk magnetocrystalline anisotropy field ($H_K$) and the second peak corresponds to the surface/interface magnetic anisotropy field ($H_{KS}$) of the system. The unique TS method of measuring magnetic anisotropy provides the advantage of measuring surface and bulk anisotropy separately. Since SSE is a surface/interface-related phenomenon, we are interested in $H_{KS}$ and its temperature evolution, as it controls surface magnetization and hence LSSE voltage. Figure 2c shows the temperature dependence of $H_{KS}$ for the YIG, YIG/Pt, and YIG/$C_{60}$/Pt samples. As expected, $H_{KS}$ ($T$) for YIG shows a peak around 75 K, which has been attributed to the rotation of surface spins away from the perpendicular easy-axis direction.[25] It is worth noting in Fig. 2c that while the coating of Pt on the surface of YIG considerably *increases* $H_{KS}$ in YIG/Pt, as compared to YIG, the coating of $C_{60}$ on the surface of YIG drastically *decreases* $H_{KS}$ in YIG/$C_{60}$/Pt. The decrease in $H_{KS}$ in YIG/$C_{60}$/Pt can be due to the hybridization between the $d_z^2$ orbital of Fe and p orbitals of C atoms.[24] As $H_{KS}$ inversely scales with $V_{\text{LSSE}}$,[25] the decrease in $H_{KS}$ of YIG due to $C_{60}$ interface is expected to increase $V_{\text{LSSE}}$ in YIG/$C_{60}$/Pt relative to YIG/Pt.

To provide an in-depth understanding of the interfacial properties of the YIG/$C_{60}$/Pt system in regards to the experimental measurements, we perform first principles simulation based on DFT for the atomic and electronic structure properties of the studied system. Details of the computational approach are given in the Methods section. Due to the limitations imposed by the complexity of the studied composite and the potentially large number of atoms in the supercell, we take advantage of the fact that the majority of the magnetic moment within the YIG unit cell is highly localized to the Fe sites[30], and thus model the YIG layer by a Fe layer. The periodically distributed $C_{60}$ molecules between Fe and Pt layers, each composed of three monolayers, are shown in Fig. 3a, where some characteristic distances are also denoted. To get a better idea of the atomic bonds of C-Pt and C-Fe, a schematic accentuating the atomic locations directly above and below of the buckyball are shown in Fig. 3b together with some relative displacements with respect to the horizontal planes fixed by the rest of the atoms in the Fe and Pt layers. It is found that, in the relaxed configuration, the $C_{60}$ molecule is chemisorbed with one hexagonal C face atop of one Pt atom, which gets pushed below the initial layer position by $\delta_{Pt}^{(1)} \approx 0.4$ Å, while the neighboring Pt atoms are pulled above by $\delta_{Pt}^{(2)} \approx 0.2$ Å. The Fe atoms in the immediate vicinity of the $C_{60}$ cage are also displaced by $\delta_{Fe} \approx 0.1$ Å forming an armchair. The C atoms in the hexagonal face adjacent to the Pt layer form chemical bonds with a length of $d_{Pt-C} \approx 2.2$ Å, while those in the hexagon close to the Fe layer form bonds with inequivalent lengths of $\delta_{Fe-C}^{(1)} \approx 2$ Å and $\delta_{Fe-C}^{(2)} \approx 2.25$ Å, respectively. The distance between the $C_{60}$ molecules is $d_{C60-C60} \approx 4.13$ Å and it is larger than the overall separation of 3.13 Å of the buckyball crystal[31], which ensures a minimal interaction between adjacent buckyballs. All structural parameters are summarized in Table 1, where results including the effects of the van der Waals interaction via the DFT-D3 approach are also shown. Our calculations indicate that most of the characteristic distances do not change significantly, although $D$ is reduced by 0.01 Å, while $\delta_{Pt}^{(1)}$ and $\delta_{Pt}^{(2)}$ are reduced by 0.04 Å and 0.03 Å, respectively, upon taking the van der Waals dispersion into account.

The calculated average magnetic moments per layer are also shown in Fig. 3b. The C atoms in the hexagonal face close to the Fe layer acquire an antiparallel average magnetic moment of about 0.01 $\mu_B$. The intercalation of the $C_{60}$ molecule decreases the magnetic moments of the interfacial Fe atoms in the immediate vicinity of the $C_{60}$ cage to 2.27 $\mu_B$, while it reduces to zero the proximity induced magnetization of the Pt atoms in the first layer near the $C_{60}$ cage. This suggests that there is a reduction of the interfacial PMA, which is consistent with our measurements of $H_{KS}$ in YIG/$C_{60}$/Pt, and it is attributed to the hybridization of the Fe $dz^2$ orbitals with C $p_z$ orbitals[24].

We also calculate the electronic density of states (DOS), where the spin-resolved results are given in Fig. 3c, while the total density of states is shown in Fig. 3d. In addition to DOS for the Pt/$C_{60}$/Fe, we also present the obtained DOS for the Pt/Fe for comparison. The Fe/Pt structure is formed by removing the $C_{60}$ molecules and allowing the adjacent layers of Pt and Fe to relax and bond. It is interesting to note that while the Pt/Fe system exhibits a spin-polarized DOS near the position of the Fermi level, with the minority spins having the dominant contribution to the transport, for the Pt/$C_{60}$/Fe structure both spins contribute almost equally to the DOS at the Fermi level (Fig. 3c). This ultimately leads to a reduction in the conductivity mismatch between the Pt and Fe layers as a result of the $C_{60}$ interface. This situation is further clarified by the total DOS in Fig. 3d, which shows a significant enhancement of the conduction states near $E_F$ for the Pt/$C_{60}$/Fe heterostructure as compared to the Pt/Fe system. In fact, it is found that DOS at $E_F$ for Pt/$C_{60}$/Fe is about 600% larger than DOS at $E_F$ for Pt/Fe, which further corroborates the giant SSE enhancement in our experiments. Even though the synthesized samples involve layers with different $C_{60}$ thicknesses, we note that the individual buckyball has an energy gap between its highest occupied molecular orbital and the lowest unoccupied molecular orbital, which is very similar to the semiconducting gaps of a linear chain of $C_{60}$ molecules[32], thus the characteristic DOS behavior is due to the interface effects with the Pt and Fe layers and they are expected to be preserved regardless of the thickness of the $C_{60}$ layer. Thus the simulated structure as depicted in Fig. 3a is expected to be a good representative of the measured samples.

**Effect of $C_{60}$ on the LSSE voltage in the YIG/$C_{60}$/Pt structure:** Figures 4a and b show the LSSE voltage ($V_{LSSE}$) *vs.* magnetic field ($H$) curves for YIG/$C_{60}$/Pt samples with varying $C_{60}$ thicknesses ($t_{C60}$ = 0, 5, 10, 30, and 50 nm) at two representative temperatures of 140 and 300 K, for a temperature gradient of $\Delta T$ = 2 K. It can be seen in this figure that in the low field region ($H \leq \pm 0.3$ kOe), $V_{LSSE}$ is relatively small (almost zero) and remains almost unchanged with increasing the magnetic field. This low field anomaly has been attributed to the presence of the surface PMA of YIG.[18,25] It is worth mentioning in the present case that even after the introduction of the $C_{60}$ layer between the YIG and Pt layers, the anomalous low field $V_{LSSE}$ ($H$) behavior is still persistent, underlining the same mechanism for LSSE voltage generation in YIG/$C_{60}$/Pt. Saturated $V_{LSSE}$ has been calculated as the average of positive and negative peak voltages. At 300 K, YIG/Pt with no $C_{60}$ layer has produced $V_{LSSE}$ of 110 nV (Fig. 4a). When a 5nm $C_{60}$ film was introduced between the YIG and Pt interfaces, $V_{LSSE}$ increased to 190 nV. The enhancement of $V_{LSSE}$ due to the $C_{60}$ intermediate layer becomes more prominent at low temperature. At 140 K, $V_{LSSE}$ increases from 70 to 660 nV with the insertion of the 5 nm $C_{60}$ thin film (Fig. 4b). As can be summarized in Fig. 4c, when the thickness of $C_{60}$ is increased from 5 to 50 nm, $V_{LSSE}$ decreases sharply first and then gradually. At 300 K, YIG/C$_{60}$/Pt samples with $t_{C60}$ = 10, 30, and 50 nm show smaller values of $V_{LSSE}$ as compared to YIG/Pt. At 140 K, however, the opposite trend is observed. For the thickest C$_{60}$ layer (50 nm), the $V_{LSEE}$ value of YIG/C$_{60}$/Pt is still greater than that of YIG/Pt.

To elucidate the observed phenomenon, we have studied in detail the temperature evolution of $V_{LSEE}$ in YIG/Pt, YIG/C$_{60}$(5 nm)/Pt, YIG/C$_{60}$(10 nm)/Pt, YIG/C$_{60}$(30 nm)/Pt, and YIG/C$_{60}$(50 nm)/Pt. All measurements were performed from 300 to 140 K, and the results are shown in Fig. 5a. It should be recalled that for YIG/Pt (with no C$_{60}$ layer) as the temperature was decreased, $V_{LSEE}$ decreased with a slope change around 170 K, and this slope change has been attributed to the effective magnetic anisotropy change in YIG, due to spin reorientation transition.[25] However, all YIG/C$_{60}$/Pt samples have shown an opposite temperature dependence of $V_{LSEE}$; $V_{LSEE}$ remains almost constant up to 200 K but below which it starts increasing exponentially (Fig. 5a). To better visualize this, the LSSE signal normalized to the signal at 140 K is shown in Fig. 5b. The exponential rise of $V_{LSEE}$ below 200 K is evident from this figure, for all C$_{60}$-coated samples. All this suggests a dominant effect of C$_{60}$ deposition on the LSSE signal in the YIG/C$_{60}$/Pt systems.

It has been experimentally shown that the spin diffusion length of C$_{60}$ possesses an exponential increase with lowering temperature just below 200 K when the film thickness is below 60 nm.[33] This logically relates the temperature dependence of $V_{LSEE}$ to that of the spin diffusion length of C$_{60}$. In other words, the strong increase of $V_{LSEE}$ with a temperature below 200 K can be attributed to the strong temperature dependence of the spin diffusion length of C$_{60}$ in this temperature region. To verify this, the $V_{LSEE}$ (*T*) data has been fitted to an exponential function that can be used to describe the temperature dependence of the spin diffusion length of $C_{60}$, and an example of this fit is shown in the inset of Fig. 5b. This result indicates that the long spin diffusion length of $C_{60}$ has indeed played a crucial role in promoting spin transport in YIG/$C_{60}$/Pt.

To quantitatively explain the $C_{60}$ thickness-dependent LSSE behavior in the YIG/$C_{60}$/Pt systems, we have adapted the model proposed for YIG/Pt by Lin *et al.*[12] Since the exchange spin transport mechanism is dominant in the organic material $C_{60}$, the spin current can be written in the following form:

$$J_s^{Pt/C60/YIG} = \frac{\left(k\nabla T e^{-\left(\frac{x}{\lambda_{Pt}}\right)}\right)}{1+G_{YIG}\left(\frac{1}{G_{\frac{Pt}{C60}}}+\frac{1}{G_{Pt}}\right)} \frac{1}{\cosh\left(\frac{t_{c60}}{\lambda_{c60}}\right)+G_{c60}\left(\frac{1}{G_{Pt}}+\frac{1}{G_{\frac{C60}{YIG}}}\right)\left(\sinh\left(\frac{t_{c60}}{\lambda_{c60}}\right)\right)},$$

(2)

where $k$ is the spin current coefficient, $\nabla T$ is the temperature gradient, $G$ is the spin current conductance, $\lambda$ is the spin diffusion length of the corresponding material, $t_{C60}$ is the thickness of the $C_{60}$ layer. Since it is difficult to obtain spin current magnitude from LSSE measurements, we have considered the ratio of the spin currents in both cases for our comparison purpose,

$$\frac{J_s^{Pt/C60/YIG}}{J_s^{Pt/YIG}} = \left(1+\frac{\left(\left(\frac{G_{\frac{Pt}{C60}}}{G_{\frac{Pt}{YIG}}}\right)-1\right)G_{Pt}}{G_{\frac{Pt}{C60}}+G_{Pt}}\right)\frac{1}{\cosh\left(\frac{t_{c60}}{\lambda_{c60}}\right)+G_{c60}\left(\frac{1}{G_{Pt}}+\frac{1}{G_{\frac{C60}{YIG}}}\right)\left(\sinh\left(\frac{t_{c60}}{\lambda_{c60}}\right)\right)}$$ (3)

In the spin-wave approximation,[12,34] $G_{\frac{Pt}{YIG}}$ is proportional to $(T/T_C)^{3/2}$, where $T_C$ is the Curie temperature of YIG. Since $T_C$ of YIG is very high (~560 K), spin current injection should increase as a thin layer $C_{60}$ is sandwiched in between YIG and Pt. This can explain the increased $V_{LSEE}$ for YIG/$C_{60}$(5nm)/Pt relative to YIG/Pt. It was previously reported that the insertion of a thin antiferromagnetic NiO layer (~1 nm) between YIG and Pt increased $V_{LSEE}$.[12] However, since $\lambda$ of NiO is relatively small (~1-2 nm), the hyperbolic function in the denominator of Eq. (3) increases, resulting in reduction of the spin current, as the NiO thickness is increased. In our case, as the thickness of the $C_{60}$ layer is increased, both hyperbolic terms in Eq. (3) increase, resulting in an exponential decrease of $V_{LSEE}$. At the same time, reduction in $H_{KS}$ in YIG/$C_{60}$/Pt due to hybridization between the $d_z^2$ orbital of Fe and C atoms, which is evident from our TS studies and DFT calculations, would result in a net increase of spin moments at the YIG surface.[24] Also, it has recently been shown that the Stoner criteria for magnetism can be beaten in $C_{60}$ by the metal-molecule interface and can induce a magnetic moment in the metal surface.[23] Theoretical studies have shown that increase in surface magnetic moment density increases the spin mixing conductance.[14] This explains our observation of the enhanced $V_{LSSE}$ at room temperature in YIG/5nm $C_{60}$/Pt for both single crystal and thin film[38] of YIG as compared to YIG/Pt, when $\lambda_{c60}$ is relatively small at room temperature (~12 nm). The decrease in $V_{LSSE}$ with increasing the $C_{60}$ thickness ($t_{C60}$) for the YIG/$C_{60}$/Pt systems can be decribed by the relation $V_{LSSE} \propto e^{-t/\lambda}$.[39] Fiting the $C_{60}$ thickness-dependent $V_{LSSE}$ data of YIG/5nm $C_{60}$/Pt at 300 K to this equation has yielded $\lambda_{c60}$ ~11±2 nm, which is similar to that reported for the $Ni_{80}Fe_{20}$/$C_{60}$/Pt system ($\lambda_{c60}$ ~13±2 nm at 300 K) using the spin pumping method.[39] This indicates that the $C_{60}$ spin current arriving at the $C_{60}$/Pt interface is proportional to $e^{-t/\lambda}$.

In conclusion, we have demonstrated a new, effective approach for enhancing the LSSE in a ferromagnetic insulator/metal system like YIG/Pt by adding a thin, intermediate layer of high spin diffusion length organic semiconductor like $C_{60}$. We have shown that the presence of $C_{60}$ reduces significantly the conductivity mismatch between YIG and Pt and the surface magnetic anisotropy of YIG, giving rise to the enhanced spin mixing conductance and hence the enhanced LSSE. Results from our first principles simulations demonstrate that the density of carriers at the Fermi level is much enhanced upon inclusion of the interface $C_{60}$ layer, which is also accompanied by a reduced magnetic anisotropy. The LSSE of YIG/$C_{60}$/Pt strongly depends on the spin diffusion length of the intermediate layer $C_{60}$; the temperature dependence of LSSE resembles that of the spin diffusion length of $C_{60}$. Our study provides a pathway for designing novel hybrid materials with prospective applications in spin caloritronics and other multifunctional devices.

**Methods**

**Sample characterization.** Single crystal YIG was purchased from Crystal Systems Corporation, Hokuto, Yamanashi, Japan, which was grown using Floating zone method along (111) direction. Various layers of $C_{60}$ were deposited on top of the YIG surface, using the thermal evaporation method with the evaporation rate of 0.2 Å/s at the base pressure of 2 x $10^{-7}$ torr. Figure 1b shows the SEM cross-section and EDX color map image of the 50 nm thick $C_{60}$ deposited YIG slab. From the SEM image, it can be concluded that the $C_{60}$ was evenly deposited on the surface of the YIG single crystal slab.

**Measurements.** Longitudinal spin Seebeck voltage measurements were performed on a YIG single crystal of dimension 6 mm × 3 mm × 1 mm (length × width × thickness). A platinum strip of 6 mm × 1 mm × 15 nm was deposited on YIG using DC sputtering. The sputtering chamber was evacuated to a base pressure of $5 \times 10^{-6}$ Torr and Argon pressure of 7 mT during the deposition. DC current and voltage used for deposition were 50 mA and ~ 350 V, respectively. The schematic of the LSEE measurement set-up is shown in Fig. 1a. For LSSE measurements

YIG/Pt was sandwiched between two copper plates. A Peltier module was attached to the bottom plate and top plate temperature was controlled through molybdenum screws attached to the cryogenic system. A temperature gradient of approximately 2 K was achieved by applying 3 A current to the Peltier module. K-type thermocouples were used to monitor the temperature of the top and bottom plates. After stepping the system temperature and Peltier module current, measurements were performed after 2 h of stabilization time. The SSE voltage was recorded as the magnetic field was swept between positive and negative saturation of YIG, using a Keithley 2182 Nano voltmeter.

Transverse susceptibility (TS) measurements were performed using a self-resonant tunnel diode oscillator with a resonant frequency of 12 MHz and sensitivity in resolving frequency shift on the order of 10 Hz.[25,26] The tunnel diode oscillator is integrated with an insert that plugs into a commercial Physical Properties Measurement System (PPMS, Quantum Design), which is used to apply dc magnetic fields (up to $\pm 7$ T) as well as provide the measurement temperature range (10 K < T < 300 K). In the experiment, the sample is placed in an inductive coil, which is part of an ultrastable, self-resonant tunnel-diode oscillator in which a perturbing small RF field ($H_{AC} \approx$ 10 Oe) is applied perpendicular to the DC field. The coil with the sample is inserted into the PPMS chamber which can be varied the temperature from 10 K to 350 K in an applied field up to 7 T.

**Computational Methods**. The first principles simulations are performed using the local spin density approximation to the density functional theory (DFT) as implemented in the Quantum ESPRESSO package [35]. We use ultrasoft pseudopotentials with a kinetic energy cutoff of 320 eV. The exchange-correlation is treated within the Perdew-Burke-Ernzerhof (PBE) generalized gradient approximation[36]. For the calculations, we construct a supercell consisting of equally spaced $C_{60}$ buckyballs sandwiched between a Fe layer composed of three bcc Fe (001) monolayers and a Pt layer composed of three fcc Pt (001) monolayers. In total the supercell consists of 156 atoms with 48 Pt atoms, 48 Fe atoms and 60 C atoms. The reciprocal space is sampled with a uniform Monkhorst-Pack 4 x 4 x 1 mesh. The outermost Fe and Pt monolayers are kept fixed during the relaxation, and all the other atoms are allowed to relax until the change in energy is less than $10^{-5}$ eV and the forces acting on atoms are less than 0.02 eV/A. We also included the vdW-D3 dispersion correction[37] in the calculations.

## Acknowledgments

Research at USF was supported by the Army Research Office through Grant No. W911NF-15-1-0626 (Spin-thermo-transport studies) and by the U.S. Department of Energy, Office of Basic Energy Sciences, Division of Materials Sciences and Engineering under Award No. DE-FG02-07ER46438 (Magnetic studies). LW acknowledges support from the US Department of Energy, Office of Basic Energy Sciences, under Grant No. DE-FG02-06ER46297. The use of the University of South Florida Research Computing facilities are also acknowledged. TN acknowledges support from the STYLENQUAZA LLC. DBA VICOSTONE USA.

## Author contributions

V.K., R.G. and R.D. had equal contributions to the work. M.H.P. and T.N. developed the initial concept. V.K., R.D., R.G., T.N., and M.H.P. designed the study. YIG/$C_{60}$ samples were fabricated by R.G. and M.P. YIG/$C_{60}$/Pt samples were fabricated by V.K. and R.D. Structural and magnetic characterization, and spin Seebeck effect measurements were performed and analyzed by V.K., and R.D. A.P. and L.M.W. performed DFT calculations and simulations. All authors discussed the results and wrote the manuscript. M.H.P. and H.S. jointly led the research project.

**Additional information**

**Competing financial interests:** The authors declare no competing financial interests.

Corresponding authors: phanm@usf.edu (M.H.P);

ngtho@uga.edu (N.D.T.); sharihar@usf.edu (H.S.)

**Figure captions**

**Figure 1** (a) Measurement geometry and spin transport through the YIG/$C_{60}$/Pt layer. Cross-sectional SEM and EDX color map images of the 50 nm thick $C_{60}$ deposited YIG slab.

**Figure 2** (a) Magnetic hysteresis (*M-H*) loops taken at 300 K for YIG, YIG/Pt, and YIG/$C_{60}$/Pt structures; (b) Transverse susceptibly spectra taken at 200 K for YIG, YIG/Pt, and YIG/$C_{60}$/Pt structures; and (c) Temperature dependence of surface/interface perpendicular magnetic anisotropy field ($H_{SK}$) for YIG, YIG/Pt, and YIG/$C_{60}$/Pt structures.

**Figure 3** (a) The Pt/$C_{60}$/Fe heterostructure, where Pt and Fe atoms are indicated with grey and red colors and C atoms are in yellow. Characteristic interatomic distances are also denoted. (b) Schematic representation of the Pt/$C_{60}$/Fe relaxed structure, showing the buckling (enhanced to help visualization) of the innermost Pt and Fe layers due to the intercalation of the $C_{60}$ molecules. Some characteristic vertical displacements are also shown. The numbers on the right represent the calculated average magnetic moments, in Bohr magneton, per layer. The considered C layers in are composed of the C hexagons closest to the metallic surfaces. (c) Spin resolved DOS for the Pt and Fe layers, with and without the $C_{60}$ molecule. (d) Total DOS for the same structures as in (c).

**Figure 4** LSSE voltage vs. magnetic field curves taken at (a) 300 K and (b) 140 K for YIG/$C_{60}$/Pt with different thicknesses of $C_{60}$.

**Figure 5** (a) $C_{60}$ thickness dependence of the LSSE signal at 300 K and 140 K; (b) Temperature dependence of LSSE voltage for YIG/$C_{60}$/Pt with different thicknesses of $C_{60}$; (c) The normalized value of LSSE for different $C_{60}$ thicknesses of YIG/$C_{60}$/Pt. Inset of (c) shows the fit for the 5 nm $C_{60}$ thickness. The temperature dependence of spin diffusion length determined from our SSE method and the MR method.[33]

**Table 1**

Relaxed structural parameters of the Fe/C60/Pt layer shown in Figs 3a and 3b. Specifically, $D$, $d_{Fe-C}^{(1,2)}$, $d_{Pt-C}$, $d_{C60-C60}$, $\delta_{Pt}^{(1,2)}$, and $\delta_{Fe}$ (in Å ) represent the separation between the Fe and Pt layers, the lengths of the covalent bonds formed between the Fe and C atoms, between Pt and C atoms, the distance between the C$_{60}$ molecules, and the displacements of the Pt and Fe atoms located immediately in the vicinity of the C$_{60}$ molecule, respectively. The numbers in parenthesis represent the same distances calculated by including the vdW correction.

| $D$ (Å) | $d_{Fe-C}^{(1)}$(Å) | $d_{Fe-C}^{(2)}$(Å) | $d_{Pt-C}$(Å) | $d_{C60-C60}$(Å) | $\delta_{Pt}^{(1)}$(Å) | $\delta_{Pt}^{(2)}$(Å) | $\delta_{Fe}$(Å) |
|---|---|---|---|---|---|---|---|
| 10.33 | 2.021 | 2.25 | 2.21 | 4.13 | 0.44 | 0.22 | 0.11 |
| (10.31) | (2.02) | (2.25) | (2.198) | (4.13) | (0.4) | (0.18) | (0.11) |

**Figure 1**

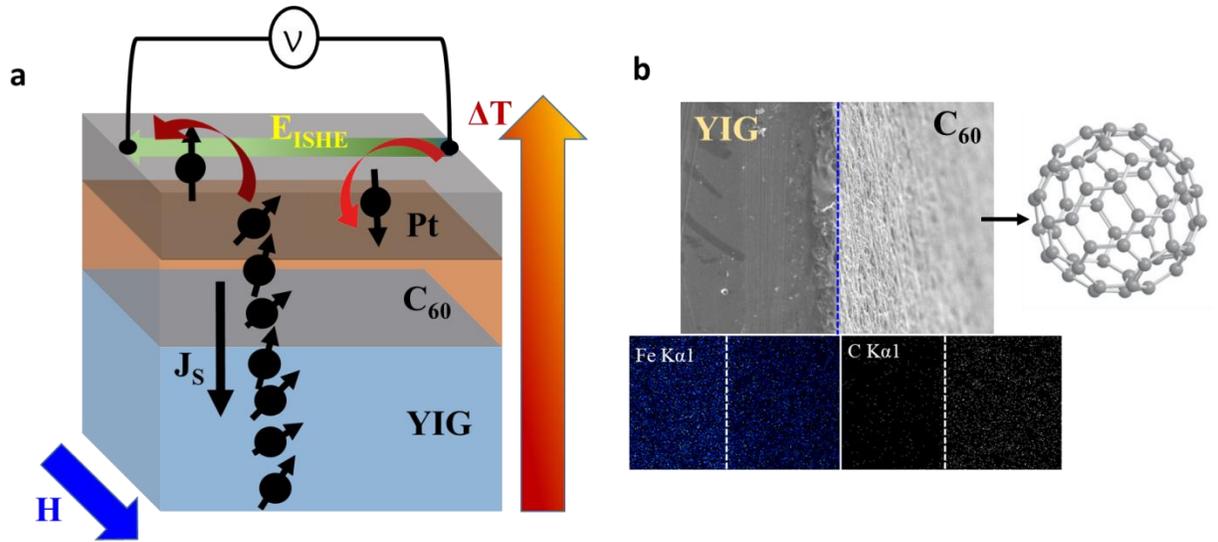

**Figure 2**

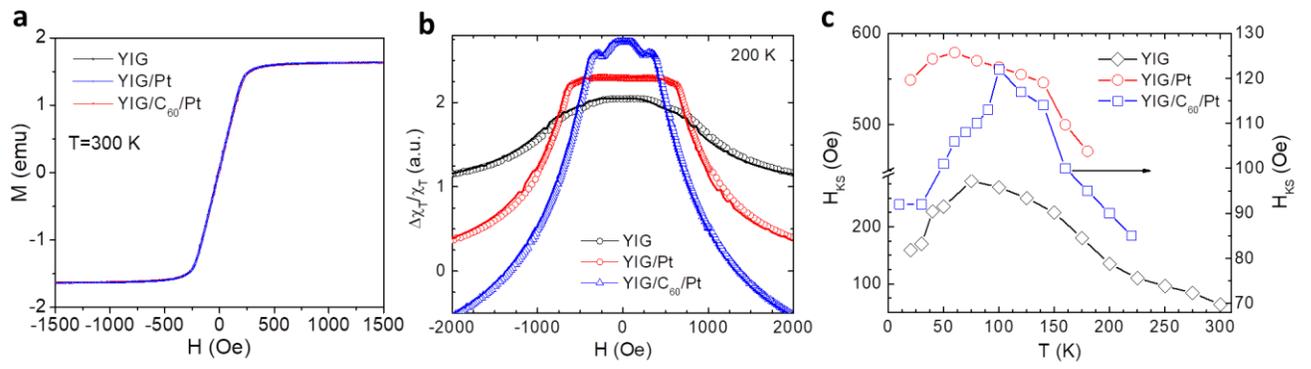

**Figure 3**

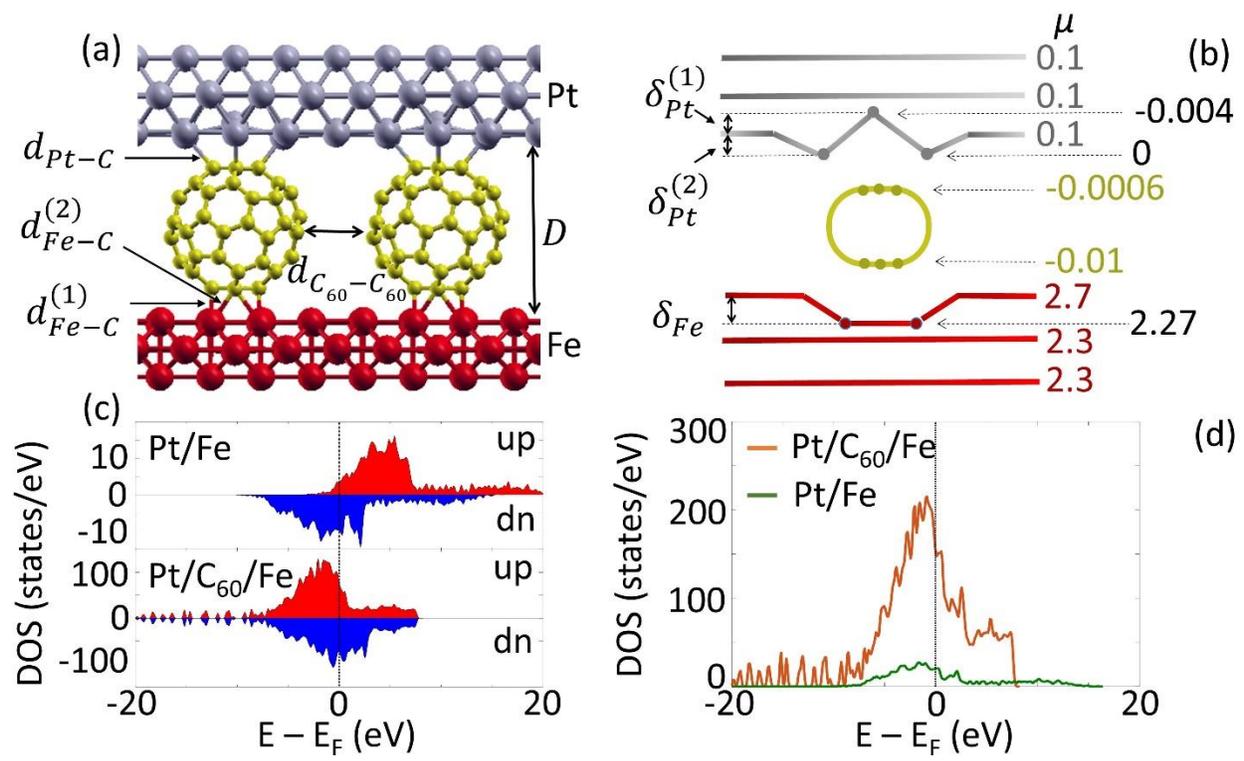

**Figure 4**

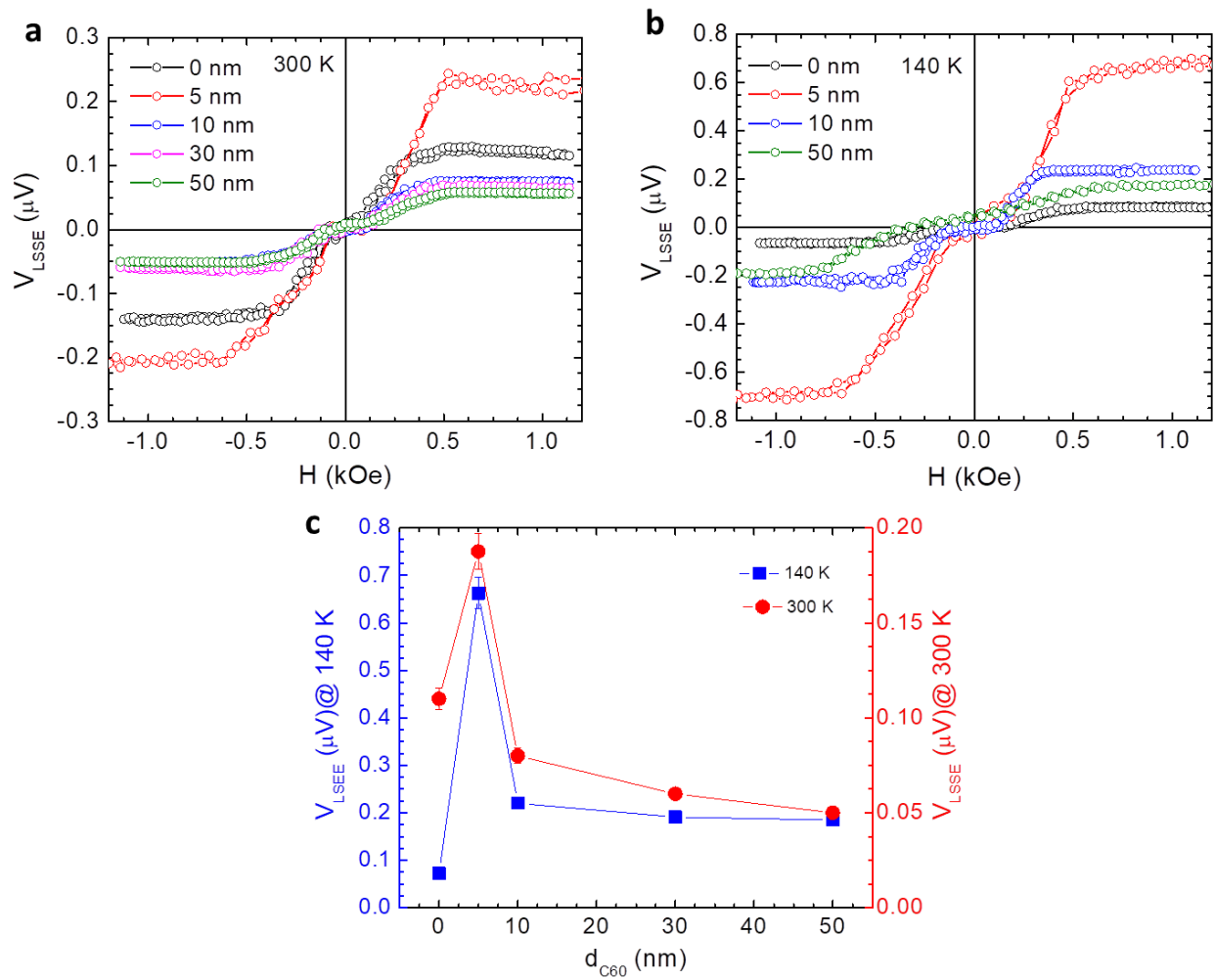

**Figure 5**

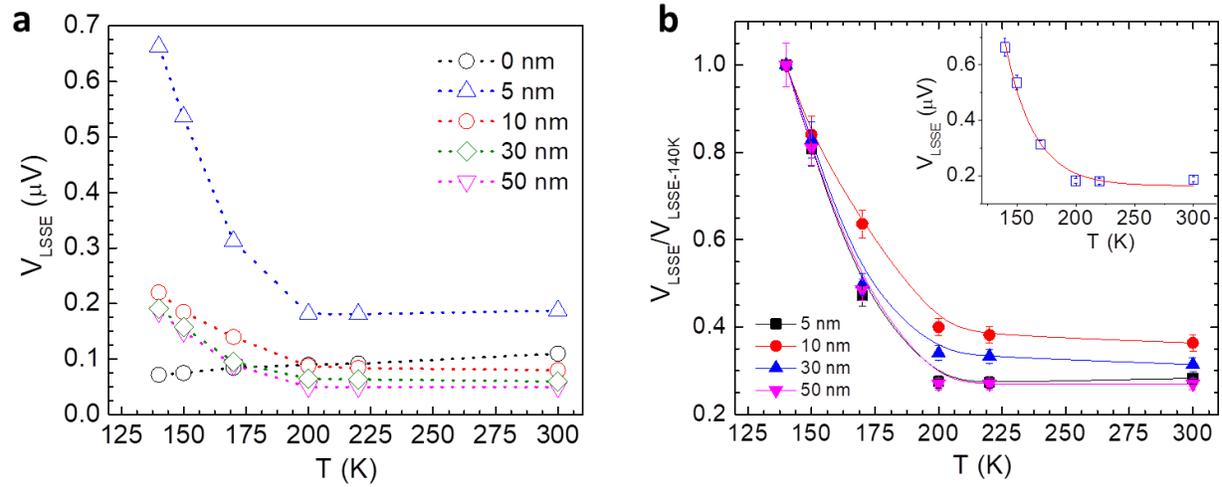